# Highly Conductive $Co_3Se_4$ Embedded in N-doped 3D Interconnected Carbonaceous Network for Enhanced Lithium and Sodium Storage

*Bingke Liu,[a] Junming Cao,[a] Junzhi Li,[a] La Li, [a*] Duo Chen,[a] Siqi Zhang,[c] Dong Cai,[a] and Wei Han,[a,b*]*

[a]Sino-Russian international joint laboratory for clean energy and energy conversion technology, College of Physics, Jilin University, Changchun City 130012, P. R. China.
E-mail: lali910217@gmail.com
[b]International Center of Future Science, Jilin University, Changchun City 130012, P. R. China.
E-mail: whan@jlu.edu.cn
[c]Key Laboratory of Physics and Technology for Advanced Batteries (Ministry of Education), College of Physics, Jilin University, Changchun 130012, P. R. China.

## Abstract

Traditional cobalt selenides as active materials in lithium-ion batteries (LIBs) and sodium-ion batteries (SIBs) would suffer from drastic volume expansions and large stacking effects, leading to a low cycling stability. In this work, we utilized a facile template method for preparing $Co_3Se_4$@N-CN (CSNC) that encapsulated $Co_3Se_4$ nanoparticles into 3D interconnected nitrogen-doped carbon network (N-CN). Satisfactorily, it possesses excellent cycling stability with enhanced lithium and sodium energy storage capacity. As an anode material in LIBs, CSNC exhibited a prominent reversible discharge performance of 1313.5 mAh $g^{-1}$ after 100 cycles at 0.1 A $g^{-1}$ and 835.6 mAh $g^{-1}$ after 500 cycles at 1.0 A $g^{-1}$. Interestingly, according to the analysis from cyclic voltammetry, the in-situ generated Se might provide extra capacity that leaded to a rising trend of capacity. When utilized as an anode in SIBs, CSNC delivered an outstanding capacity of 448.7 mAh $g^{-1}$ after 100 cycles at 0.1 A $g^{-1}$ and could retain 328.9 mAh $g^{-1}$ (77.2% of that of 0.1 A $g^{-1}$) even at a high current density of 5.0 A $g^{-1}$. The results demonstrate that CSNC is a superior anode material in LIBs and SIBs with great promise. More importantly, this strategy opens up an effective avenue for the design of transition metal selenide/carbonaceous composites for advanced battery storage systems.

## Introduction

For the past few decades, people have constantly explored novel advanced electrode materials, in order to reduce the gap between demand and supply of energy storage. The carbon-based functional materials play a pivotal role in the field of electrochemical energy storage due to their low cost and abundant resource [1-4]. Among various dimensional existence of carbon, such as 0D carbon nanodots, 1D carbon nanotubes and 2D reduced graphene oxide. 3D carbon nano-networks are one of the most potential system of electrode materials [5-7]. Generally, 3D carbon materials endow large specific surface area, stable structure, bringing about ultra-high electrochemical performance in alkaline metal battery applications, especially lithium-ion batteries (LIBs) and sodium-ion batteries (SIBs). According to the previous reports, a template-free nitrogen-doped 3D carbon can possess 596.1 mAh $g^{-1}$ after 1000 cycles at 1.0 A $g^{-1}$, remain 304.3 mAh $g^{-1}$ after 2000 cycles at 5.0 A $g^{-1}$ of LIBs. A spontaneous growth of 3D carbon exhibits 175.0 mAh $g^{-1}$ after 1000 cycles at 0.5 A $g^{-1}$, and capacity of 99.0 mAh $g^{-1}$ is maintained after 10000 cycles at 10.0 A $g^{-1}$ in the sodium storage [7, 8]. From the aspect of the micromorphology, 3D carbon materials are naturally gifted with various porous nanostructure, depending on the unique pore-forming process and mechanisms. As an ideal etchant, NaCl possesses the advantages of low cost, accessibility and high crystallinity, that is to say, NaCl could be regarded as an efficient pore-forming substance [5, 9, 10].

However, the major drawback limiting the widely practical application of carbon materials is relative lower theoretical capacity of 372 mAh $g^{-1}$ [11]. Considering this obstacle, many researchers dedicate novel hybrid nanomaterials to further enhance the capacity of carbon-based electrodes. Compared with oxides and sulfides counterparts, transition metal selenides endows matched electronegativity, high ionic conductivity with alkali metal ions, as well as less weak polarization phenomena, which is benefit for enhancing the Lithium and Sodium storage ability [12-15]. Cobalt selenide, which has many forms due to different valences, such as $Co_3Se_4$, $CoSe_2$, CoSe etc [16-20]. However, the application of $Co_3Se_4$ as the anode of LIBs and SIBs has been barely investigated systematically. Considering that the volume changes resulting from the lithium / sodium ions intercalation and deintercalation during charging and discharging, a three-dimensional interconnected carbon network could be utilized as a highly conductive skeleton to confine the volume deformation. Benefitting from the synergistic effect between two different materials, the composite could receive higher specific surface area and numerous ion transport channels, with a high conductivity and further improve for the electrochemical reversible capacity.

In our work, utilizing a facile template method, we use the NaCl as a pore-forming agent and glucose as a carbon source to successfully prepare ultrathin 3D interconnected nitrogen-doped carbon network (N-CN). By mixing as prepared N-CN with various concentration of cobalt contained solution, following by in-situ selenization, $Co_3Se_4$@N-CN (CSNC) electrodes was obtained. A reversible lithium storage capacity as high as 1313.5 mAh $g^{-1}$ could be achieved at the current density of 0.1A $g^{-1}$. During the cycling, a large increase in capacity, and this unusual phenomenon is reasonably explained by in-situ selenium precipitation reaction. Meanwhile, the hybrid electrode also maintains a high reversible capacity of 448.7 mAh $g^{-1}$ at 0.1A $g^{-1}$ after 100 cycles, with an excellent rate capability of over 77% retention as the current densities go up to 5 A $g^{-1}$, applying as the anode for sodium ion batteries.

## Experimental section

*Synthesis of 3D Cross-linked N-CN*

Firstly, 50 g NaCl and 3 g glucose powder were dissolved in 150 ml deionized water to form a transparent solution under continuous stirring for 2 h. Then, the resulting mixed solution was directly freeze-dried in vacuum, and the obtained mixture was ground with 3 g urea in an agate mortar. After that, the sample was heated at 700 °C for 3 h in a tube furnace under Ar atmosphere with the heating rate of 3 °C/min. The mixed powder was cooled down naturally and washed several times by deionized water to remove the NaCl and dried at 60 °C in the air overnight.

*Synthesis of $Co_3Se_4$@N-CN:*

For $Co_3Se_4$@N-CN composite, 0.1 g of the as-obtained N-CN powder was added into 30 ml solutions containing various concentration of $Co(NO_3)_2$ (0.05 M, 0.1 M, 0.2 M, and 0.5 M), respectively, all the solutions were stirred for 12 h at 45 °C. Then, the mixture was filtered, and dried at 60 °C for 12 h in the air. During annealing process, the obtained mixture was placed in the downstream side and 0.4g selenium powder was placed in the upstream side of the porcelain boat and heated at 650 °C for 2 h in a tube furnace under nitrogen atmosphere. The $Co_3Se_4$@N-CN samples were labeled as CSNC-0.05, CSNC-0.1, CSNC-0.2, CSNC-0.5 by adjusting the concentration of $Co(NO_3)_2$ (0.05 M, 0.1 M, 0.2 M and 0.5 M).

*Material Characterization*

X-ray diffraction (XRD, Japan Rigaku 2550) with Cu K_α radiation (λ=1.54056 Å) operating at 40 kV and 300 mA is employed to investigate the phases and crystalline structures. The microstructure and morphology of samples are analyzed by Transmission electron microscope (TEM, JEOL JSM-2010F) and Scanning electron microscope (SEM, Magellan 400). The specific surface area and the pore size distribution are measured by a Brunauer-Emmett-Teller (BET, JW-BK 132F) analyzer. And a Thermo escalab 250Xi electron spectrometer is used to conduct the X-ray photoelectron spectroscopy (XPS) spectra.

*Electrochemical measurements*

The electrochemical performance of CSNC was tested using 2032 coin-type cell. The working electrode comprised CSNC (the active material), acetylene black (the conductivity agent), and polyvinylidene fluoride (PVDF) (the binder) with a weight ratio of 7:2:1. The slurry was coated on copper foil and dried under vacuum at 80 °C overnight. And there was about 1.5 mg mass loading of active material on each electrode. The electrolyte was a 1.0 M solution of $LiPF_6$ in ethylene carbon (EC), dimethyl carbonate (DMC) and ethylmethy carbonate (EMC) (1 : 1 : 1, v/v/v) and a 1.0 M solution of $NaClO_4$ in ethylene carbon (EC), diethy carbonate (DEC) (1 : 1, v/v).

## Results and discussion

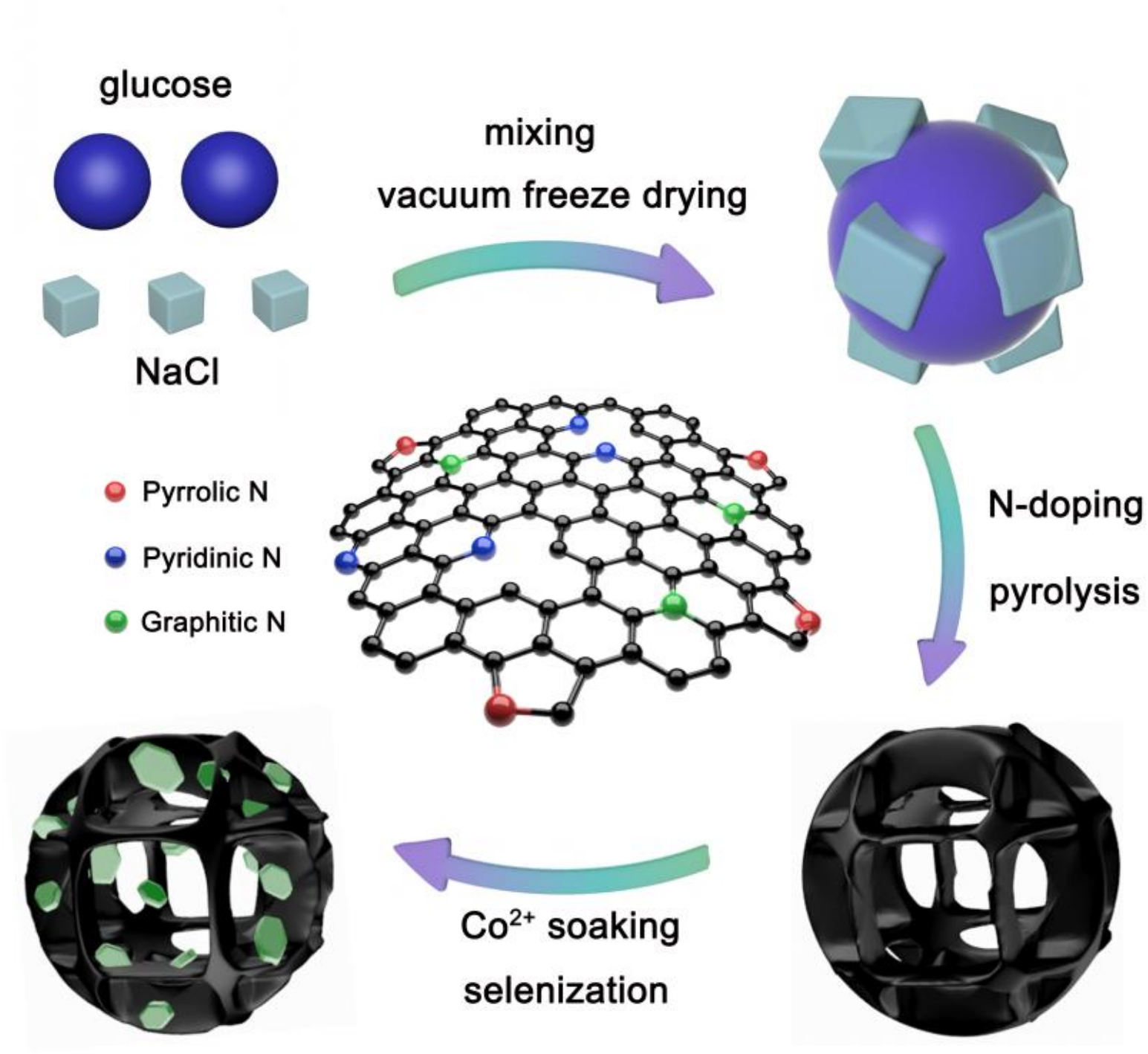

**Scheme 1.** Schematic illustration of synthesis procedure for the $Co_3Se_4$@N-CN

Scheme 1 illustrates the as-synthesized process of CSNCs. Firstly, a uniformly distributed hybrid precursor consisted of glucose and sodium chloride crystals was obtained by a simple dissolution and vacuum freeze-drying routine. Then the precursor was ground with urea and annealed in Ar atmosphere and washed to obtain the conductive N-CN. Finally, the N-CN was adequately soaked in various concentrations of $Co^{2+}$ solutions and dried overnight, then the $Co^{2+}$@N-CN undergone selenization in a tube furnace with the assistance of the selenium powder to form CSNC nanocomposite.

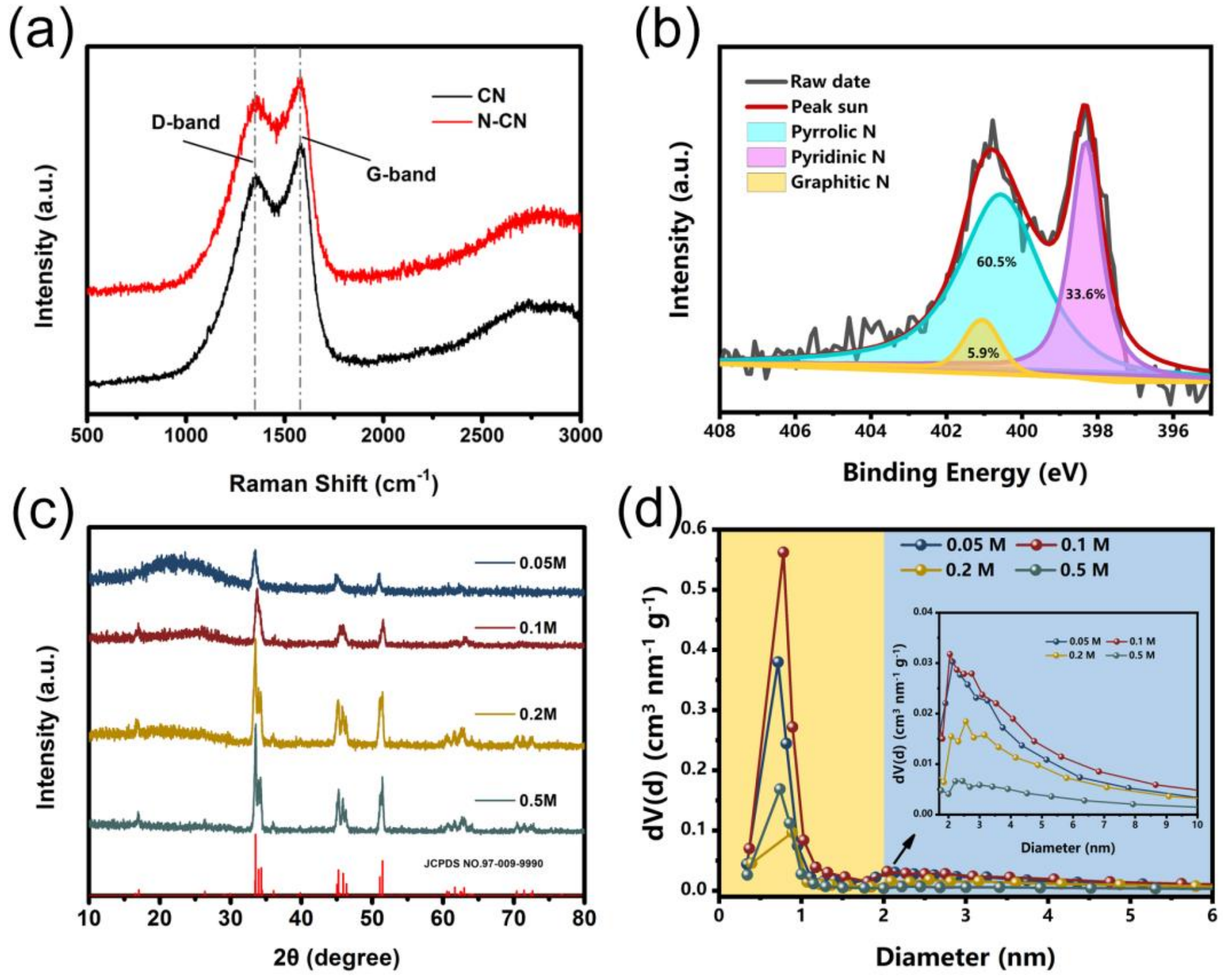

**Figure 1.** a) Raman spectra of CN and N-CN. b) XPS N 1s spectrum of the CSNC-0.1. c) XRD patterns and d) pore size distribution of CSNC-0.05, CSNC-0.1, CSNC-0.2 and CSNC-0.5.

The Raman spectra in Figure 1a shows two characteristic peaks at 1360 $cm^{-1}$ (D-band) and 1582 $cm^{-1}$ (G-band) of N-CN, which slightly shift from the peaks at 1354 $cm^{-1}$ and 1584 $cm^{-1}$ of CN on account of N-doping effect [21, 22]. The intensity ratio $I_D/I_G$ of D and G bands can be used to determine the crystallinity of carbon-based materials. Specifically, D-band is associated with edge defects, while G-band is related to highly ordered graphite [23]. The $I_D/I_G$ values of CN and N-CN are 0.88 and 0.94, respectively, which indicates the N-CN becomes more chaotic and higher degree of defect than CN, as a result of the successful introduction of N atoms. To further investigate the chemical states and surface of CSNC, the X-ray photoelectron spectroscopy (XPS) measurement was carried out. Figure S1a displays the full survey scan spectrum of C, N, Co and Se elements. As shown in Figure S1b, the C 1s spectrum can be spilt into two peaks centered at about 284.7 and 286.2 eV, respectively, corresponds to $sp^2C$-$sp^2C$ and N-$sp^3C$ bonds [24]. The deconvolution spectrum of Co 2p reveals the peaks at 780.7, 796.3, 782.3 and 797.6 eV, which can be distributed to $Co^{3+}$ $2p_{3/2}$, $Co^{3+}$ $2p_{1/2}$, $Co^{2+}$ $2p_{3/2}$ and $Co^{2+}$ $2p_{1/2}$, and the peaks at 804.2 and 786.6 eV are ascribed to the satellite peaks of Co $2p_{3/2}$ (Figure S1c) [25, 26]. The Se 3d spectrum is accurately analyzed from Figure S1d, the peaks of Se 3d are located at 55.7, 56.5, 59.3 and 58.5 eV, respectively, matching Se $3d_{3/2}$ , Se-Se , Se-O-Se and the satellite peak of Se 3d [27, 28]. In the case of N 1s, the peaks in Figure 2b are located at 398.2, 400.5 and 401.2 eV, and the peaks are assigned to Pyridinic N, Pyrrolic N and Graphitic

N, respectively. While the peak at 400.5 eV is associated to the Co-N bonding, which is caused by $Co^{2+}$ coordinating with the N-content functional group in the carbon matrix [29]. It proves that large amounts of metal ions are adsorbed on N-doped carbon. The structure of the CSNCs was further investigated by the X-ray diffraction (XRD) technique. Figure 1c shows the XRD patterns of the as-prepared CSNC samples with different $Co_3Se_4$ ratios. The characteristic diffraction peaks of all samples are quite consistent with the $Co_3Se_4$ crystal (JCPDS no. 97-009-9990), indicating that the $Co_3Se_4$ samples were successfully embedded into N-CNs. It is worth to mention that as the concentration of the $Co(NO_3)_2$ solution increases, the $Co_3Se_4$ diffraction peaks become more pronounced. This is because the particle size and the distribution density of $Co_3Se_4$ on the surface of conductive carbon network are gradually increased. And this guess is confirmed by scanning electron microscope (SEM) section as shown in the following. Besides, the XRD pattern of N-CN is displayed in Figure S2, and only a broad peak at around 22.3° can be observed, which confirms that the NaCl is completely eliminated from N-CN. The Brunauer-Emmett-Teller (BET) specific surface areas and pore size distribution of different CSNC are shown in Figure S3 and Figure 2d. The calculated BET surface areas of CSNC-0.05, CSNC-0.1, CSNC-0.2 and CSNC-0.5 are 288, 428, 161 and 150 $m^2$ $g^{-1}$, respectively (Figure S3). The CSNC-0.1 has the largest specific surface area mainly because of the appropriate pore size distribution of the N-CN network surface without blocking the mesopores. As for the CSNC-0.05 sample, the size of $Co_3Se_4$ particles is too small to contribute to the roughness on the surface of the carbon network. On the contrary, the larger sized $Co_3Se_4$ crystals of CSNC-0.2 and CSNC-0.5 with higher distributive density impede the mesopores of conductive network and further decrease the specific areas of CSNCs. This inference can be confirmed by both XRD and SEM analyses. Figure 2d provides the BET results of CSNC. Furthermore, the mesopores can enhance the interfacial contact, the micropores can provide reactive active sites [30, 31].

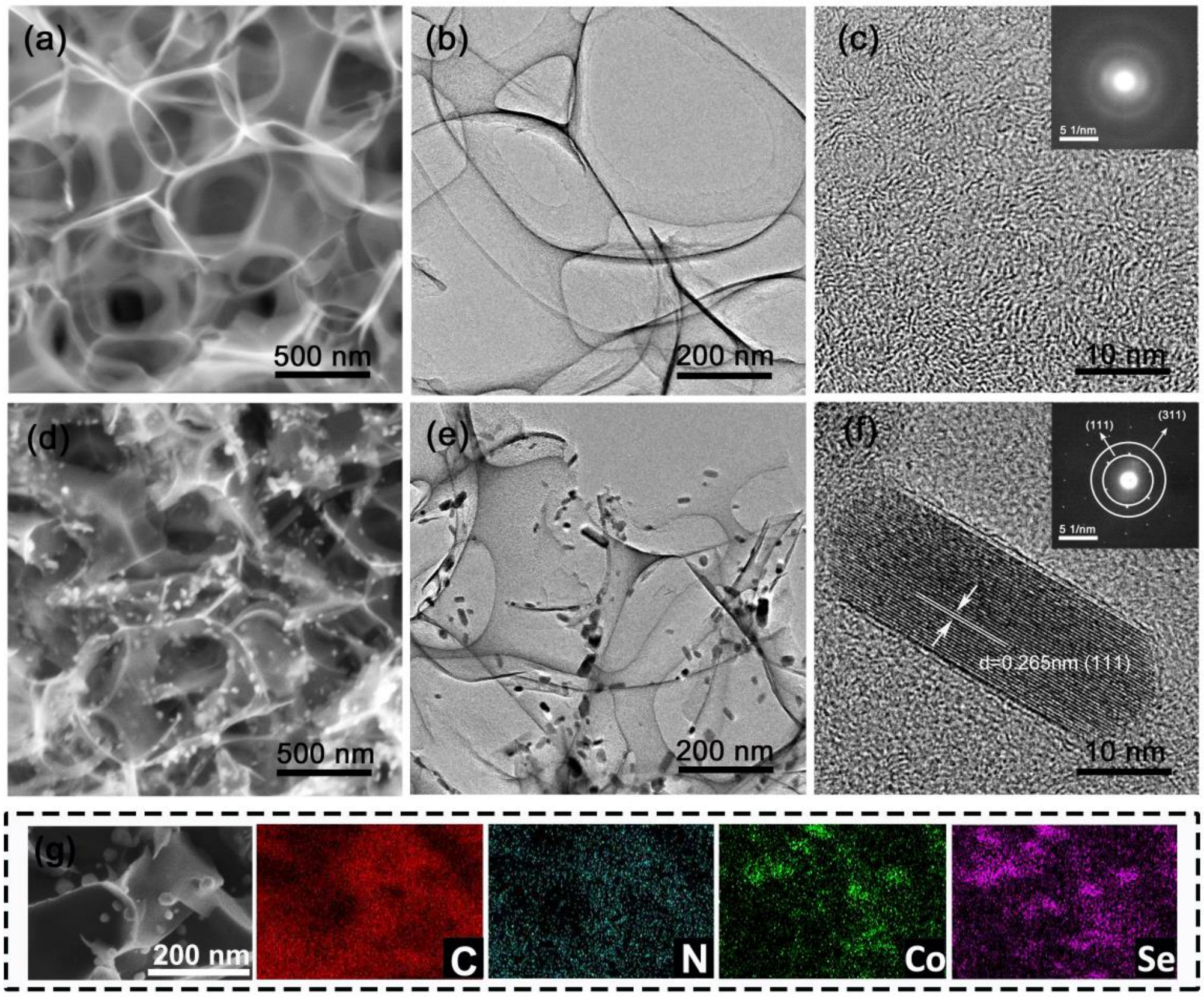

**Figure 2.** a) SEM of N-CN. b) TEM of N-CN. c) HTEM images, SAED images of N-CN. d) SEM of CSNC-0.1. e) TEM of CSNC-0.1. f) HRTEM images, SAED images of CSNC-0.1. g) Element mapping of CSNC-0.1.

In order to observe the morphology of the as-synthesized N-CN and CSNC, the SEM and transmission electron microscope (TEM) were characterized. For the N-CN, it can be observed from Figure 2a and Figure 2b that the pores caused by cubic sodium chloride crystal template possess a pore diameter from 400 to 800 nm. And the high-resolution TEM (HRTEM), selected area electron diffractions (SAED) images of N-CN are shown in the Figure 2c. The lattice fringes of N-CN have no regular pattern, indicating that N-CN is a kind of amorphous carbon. The CSNC-0.1 is also characterized by SEM and TEM measurements, as shown in Figure 2d and Figure 2e. The $Co_3Se_4$ nanoparticles uniformly distribute on N-CN with uniform particle size from about 20 nm to 40 nm. And to learn more about CSNC, the HTEM and SAED images of CSNC-0.1 are displayed in Figure 2f. The lattice spacing of $Co_3Se_4$ particles can be measured to be 0.265 nm, which is consistent with the (111) crystal planes of $Co_3Se_4$. The SAED image reveals the (111) and (311) crystal planes of $Co_3Se_4$. It proves that the $Co_3Se_4$ nanocrystals were synthesized successfully. Furthermore, the elemental mapping presented in Figure 2g indicates the C, N, Co and Se element are coexistence and uniformly distributed over the obtained CSNC-0.1. Figure S4 provides SEM images of the CSNC-0.05, CSNC-0.2 and CSNC-0.5. Contrasted with CSNC-0.1, these samples have a similar carbon network substrate, but have different sizes and distribution of $Co_3Se_4$ nanoparticles on the N-CN. The $Co_3Se_4$ particles with smallest size sparsely distribute over the CSNC-0.05 (Figure S4a), the $Co_3Se_4$ particles on the CSNC-0.2 are stacked with large size (Figure S4b), and serious stacked distribution and largest size $Co_3Se_4$ particles can be observed on the CSNC-0.5 (Figure S4c). Therefore, it is easy to ascertain that the regular change of distribution and size $Co_3Se_4$ particles can be caused by the only variable of the experiment : the concentration of $Co^{2+}$ solution. The positive correlation between the concentration of $Co(NO_3)_2$ solution and the particle size of the synthesized $Co_3Se_4$ particles was found.

**CSNCs for lithium-ion batteries**

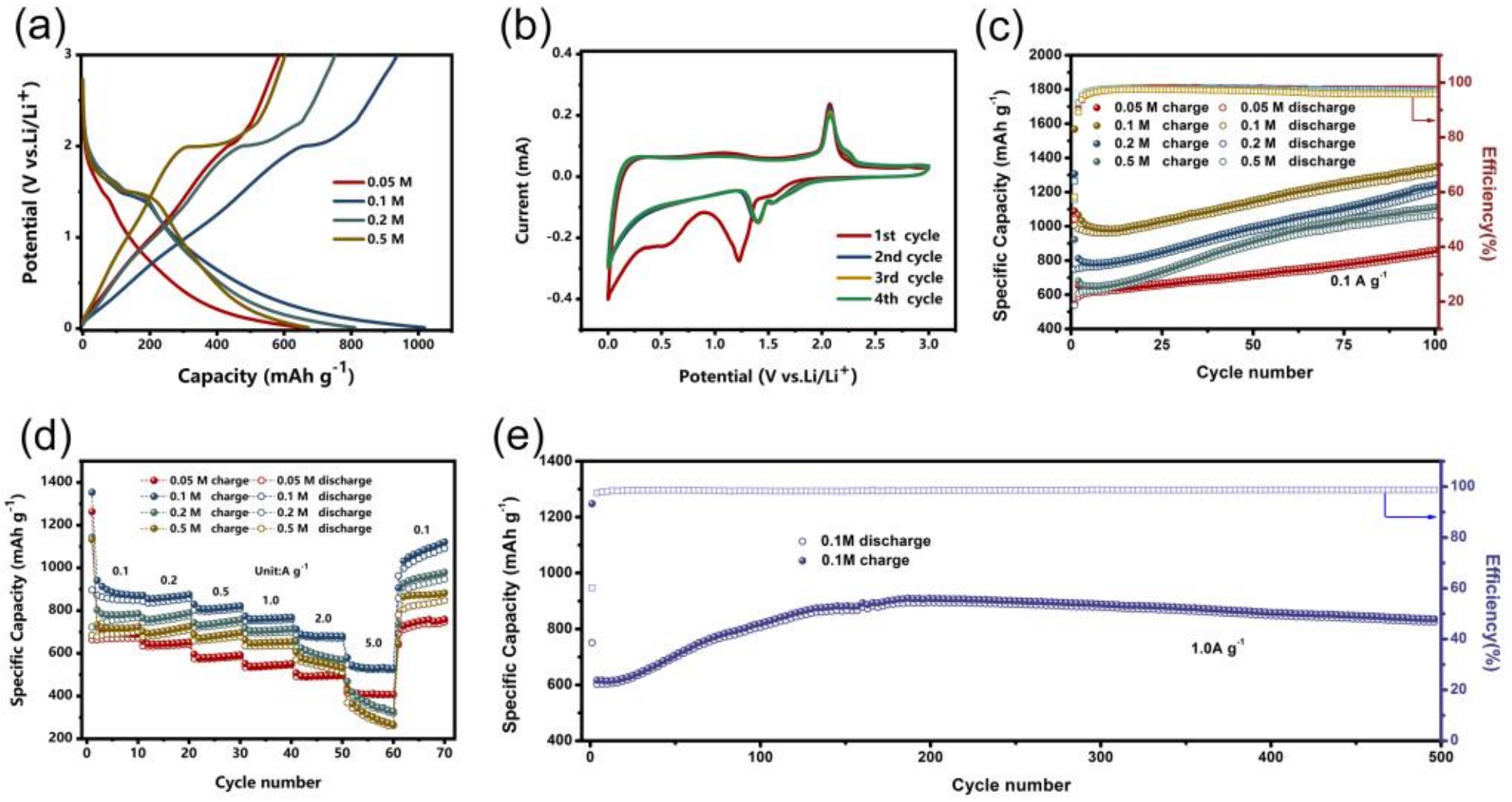

**Figure 3.** Electrochemical performance of CSNC electrode versus $Li/Li^+$. a) Galvanostatic charge/discharge curves of CSNC-0.05, CSNC-0.1, CSNC-0.2 and CSNC-0.5 at the current rate of 0.1 A $g^{-1}$. b) CV curves of CSNC-0.1 at a sweep rate of 0.1 mV $s^{-1}$. c) Cycle performance of CSNC-0.05, CSNC-0.1, CSNC-0.2 and CSNC-0.5 at the current density of 0.1 A $g^{-1}$. d) Rate performance of CSNC-0.05, CSNC-0.1, CSNC-0.2 and CSNC-0.5. e) Cycle performance of CSNC-0.1 at the current density of 1.0 A $g^{-1}$.

Electrochemical performance of CSNC electrode versus $Li/Li^+$ is depicted in Figure 3 and Figure S5. Figure S5a shows the cyclic voltammetry (CV) curves for CSNC-0.05, CSNC-0.1, CSNC-0.2 and CSNC-0.5 electrodes at a scan rate of 0.1 mV $s^{-1}$ with a cut-off voltage range of 0.01-3.0 V at room temperature. Each CSNC has significant reduction/oxidation peaks at 1.35 V, 2.10 V of the second cycle, which can be correlated with the characteristic peaks of cobalt selenide [32]. And it is found that as the concentration of $Co^{2+}$ solution increases, the area of reduction/oxidation peak increases, and the area at the low potential decreases. The reason is as the particle size of the $Co_3Se_4$ enlarges, and the distribution of $Co_3Se_4$ on N-CN is more dense, the $Co_3Se_4$ will take more participate in electrochemical reaction, thereby reduce the capacity contribution proportion of carbon. Figure 3a illustrates the constant current charge and discharge of CSNC at a current density of 0.1 A $g^{-1}$. The voltage of charge and discharge platforms can be well matched with the reduction/oxidation peaks in Figure S5a. Among all CSNCs, the CSNC-0.1 possesses the highest specific capacity and the discharge specific capacity reaches 1018.2 mAh $g^{-1}$. To learn more about the electrochemical properties of CSNC-0.1, the initial four turns of CV curves of CSNC-0.1 at the scan rate of 0.1 mV $s^{-1}$ are shown in Figure 3b. During the first lithiation process, an intense cathodic peak around 1.21 V can be related with the conversion reaction of the irreversible solid electrolyte interphase (SEI) [33]. The potential at the 1.21 V of the cathodic peaks shifts to a higher position of 1.35 V from the second cycle, and the CV curves of the 2-4 circles almost coincide, indicating that the CSNC is an excellent electrode material with good cycle stability. Figure 3c exhibits the cyclic performance of CSNC-0.05, CSNC-0.1, CSNC-0.2 and CSNC-0.5 at a current density of 0.1 A $g^{-1}$, their capacities are respectively 841.4, 1313.5, 1203.0 and 1065.0 mAh $g^{-1}$ at the $100^{th}$ cycle. The capacity of CSNC-0.1 is much higher than that of other samples during 100 cycles. This is attributed to the appropriate load, and the optimum particle size of $Co_3Se_4$ particles, which allow the N-CN and $Co_3Se_4$ nanoparticle to exert maximum synergy. In order to further research the interfacial properties of CSNC, electrochemical impedance spectroscopy (EIS) is conducted and displayed in Figure S5b. The fresh cell EIS curve of CSNC shown in Figure S5b can be divided into two parts: (1) the high frequency semicircles associated with the active materials/electrolyte resistance; (2) the low frequency diagonal corresponds to Warburg impedance related to $Li^+$ diffusion, and it can be well fitted by a classic LIBs equivalent circuit (the inset of Figure S5b). The resistance value at high frequency increases as the load of $Co_3Se_4$ on N-CN increases, this phenomenon is caused by the oversized $Co_3Se_4$ particles which will block charge transfer between the N-CN and electrolyte. It indicates that the different $Co^{2+}$ concentrations loaded on the N-CN can change the impedance and then impact the electrochemical performance of the LIB regularly. Figure 3d enumerates a comparison of the ratio performance of four kinds of CSNCs at 0.1, 0.2, 0.5, 1.0, 2.0 and 5.0 A $g^{-1}$, respectively. The good rate performance of each CSNC is due to the high specific surface area of N-CN, which provides a large number of active sites for reacting with $Li^+$. For the CSNC-0.1, its capacity can reach 941.9, 855.3, 809.8, 760.6, 689.3, 544.5 mAh $g^{-1}$ at current density of 0.1, 0.2, 0.5, 1.0, 2.0 and 5.0 A $g^{-1}$, respectively. It proves that the suitable loading ratio of $Co_3Se_4$ can provide higher energy storage than other CSNC samples at different current densities. In view of the good electrochemical performance of CSNC-0.1, the cycle performance of CSNC-0.1 at 1.0 A $g^{-1}$ is shown in Figure 3e. The capacity of CSNC-0.1 rises at the first 200 cycles and then decreases gradually. The slight capacity decrease is caused by the reduction of selenium generated in situ and the damage to the carbon skeleton structure. After 500 cycles, there is still 835.6 mAh $g^{-1}$ capacity reserved, indicates that the material can maintain a high capacity after a long period cycle at a large current density. As a comparison, the cycle performance and rate performance of N-CN are displayed in Figure S6. The N-CN has a good electrochemical performance with the capacity of 729.9 mAh $g^{-1}$ at the current density of 0.1 A $g^{-1}$ after 70 cycles (Figure S6a), and 442.2 mAh $g^{-1}$ at 1.0 A $g^{-1}$ (Figure S6b). It demonstrates that the unique 3D interconnect structure of N-CN provides a great contribution to the excellent electrochemical performance of the composite CSNC.

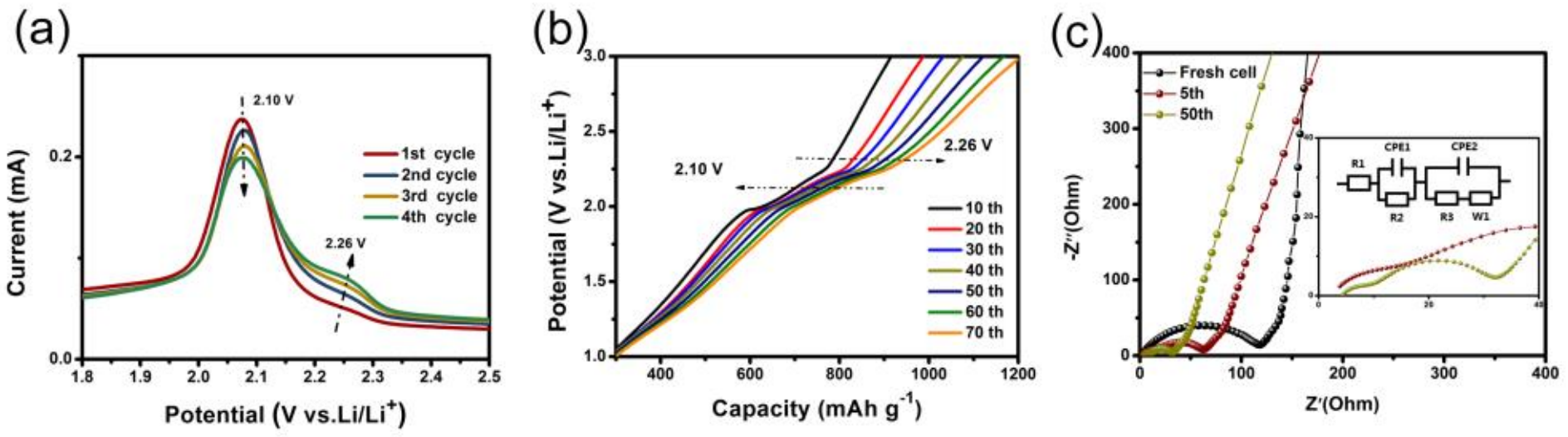

**Figure 4.** a) Enlarged details of CSNC-0.1 CV curves at a sweep rate of 0.1 mV $s^{-1}$. b) Galvanostatic charge of CSNC-0.1 at the current rate of 0.1 A $g^{-1}$. c) The EIS of CSNC-0.1 of $1^{st}$, $5^{th}$ and $50^{th}$ cycles, fitting circuit model is depicted in inset.

It is easy to detect that the capacity of four CSNCs shows a rising trend in the first 100 cycles. To learn more about the effect of this phenomenon in the reaction of LIB, the partial enlargement of the CSNC-0.1 CV curves is shown in Figure 4a. It can be seen that, the peaks at 2.26 V are gradually increased with the increasing cycles, while the reduction/oxidation peaks of $Co_3Se_4$ at 2.10 V are reduced. According to previous reports, the peaks at 2.26 V in the Figure 4a is associated with the in-situ generated Se [34]. The interesting electrochemical behavior is attributed to the in-situ generated selenide from the $Co_3Se_4$ during the charge and discharge process. It indicates that Se is constantly generated from $Co_3Se_4$, takes additional contribution of capacity in the reaction, and makes the capacity rising. The Se is reduced to $Li_2Se_n$ (n>4) for the discharge process, $Li_2Se_n$ is oxidized into insoluble $Li_2Se_n$ (n≥4) and Se during the charge process [35]. The galvanostatic charge shown in Figure 4b with the same variation trend of the 2.10 and 2.26 V, can also verify the view above. The subsequent EIS (the $5^{th}$ and $50^{th}$ cycles of CSNC-0.1 for LIBs) are depicted with a fitting equivalent circuit modeling as shown in Figure 4c. The decreased resistance during cycling indicates that the newly generated byproducts can enhance electronic conductivity. And one more semicircular impedance emerges in the high frequency of the EIS curves, revealing the appearance of new interfacial resistance [36]. The EIS curve with two semicircles can be well fitted by equivalent circuit (the inset of Figure 4c), in which the (CPE1, R2) can describe the interfacial resistance between N-CN and electrolyte, the (CPE2, R3) indicates the interfacial resistance of $Co_3Se_4$ particles and electrolyte in the internal structure of N-CN, and the interface is composed of Se, Se-O [37].

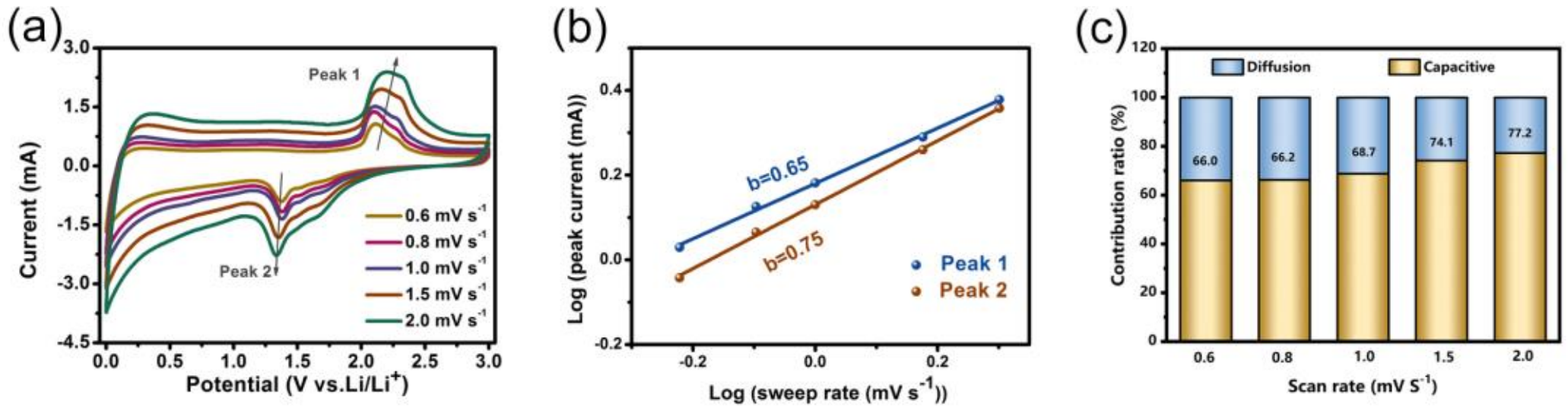

**Figure 5.** a) CV curves at different scan rates ranging from 0.2 to 2 mV $s^{-1}$. b) Log (Peak current) versus log (scan rate) at different scan rates. c) Contribution ratio of capacitive and diffusion-controlled capacities at different scan rates.

To reveal the possible reason for the superior cycling capability and excellent rate performance of the CSNC-0.1 electrode of the LIBs, detailed kinetic property was studied by different scan rates (0.6, 0.8, 1.0, 1.5 and 2.0 mV $s^{-1}$) of CSNC-0.1 CV curves shown in Figure 5a. It can be seen that different curves have similar shapes with the increased current intensity as the scan rate increases, which confirms the stable electrochemical performances. Normally the electrochemical energy storage is contributed by several different processes including surface, near-surface intercalation and pseudo-capacitance behaviors and it can be qualitatively deduced by the CV data obtained at various scan rates based on the following formula [38] :

$$i = av^b \tag{1}$$

$$\log(i) = b\log(v) + \log(a) \tag{2}$$

where $i$ is the measured current density, $v$ stands for the scan rate, and b is a constant. Generally, value of 0.5 discloses a total diffusion-controlled charge storage behaviors, while $b$ value of 1 reveals that it is the pseudocapacitive behavior dominates the charge storage process. The $b$ value can be determined by the slope of plotting log $i$ versus log $v$. As shown in Figure 5b, the $b$ value is calculated to be 0.65 of the anodic process (peak 1), 0.75 of the cathodic process (peak 2). What is more, the peaks at about 2.26 V, related to the in situ generated selenium, also slight shifts. It suggests that both capacitive behavior and diffusion-controlled processes exist in the total capacity storage of LIBs [39, 40].

Furthermore, the contribution ratio of capacitive behavior at different scan rate can be calculated by dividing between the diffusion-controlled contribution and the capacitive contribution in the light of the following equation:

$$i_{(V)} = k_1 v + k_2 v^{1/2} \tag{3}$$

where $i_{(V)}$ stands for the current density at different potential, $v$ represents the scan rate, and $k_1$ and $k_2$ are constants at a given potential. The results of the diffusion-controlled and capacitive contributions of LIB are calculated and displayed in Figure 5c. The capacitive contributions reach 68.7% and 77.2%, as the scan rates are 1.0 and 2.0 mV $s^{-1}$. A high proportion of capacitance contributes to the improvement of rate performance and long term cycling stability [41].

**CSNCs for sodium-ion batteries**

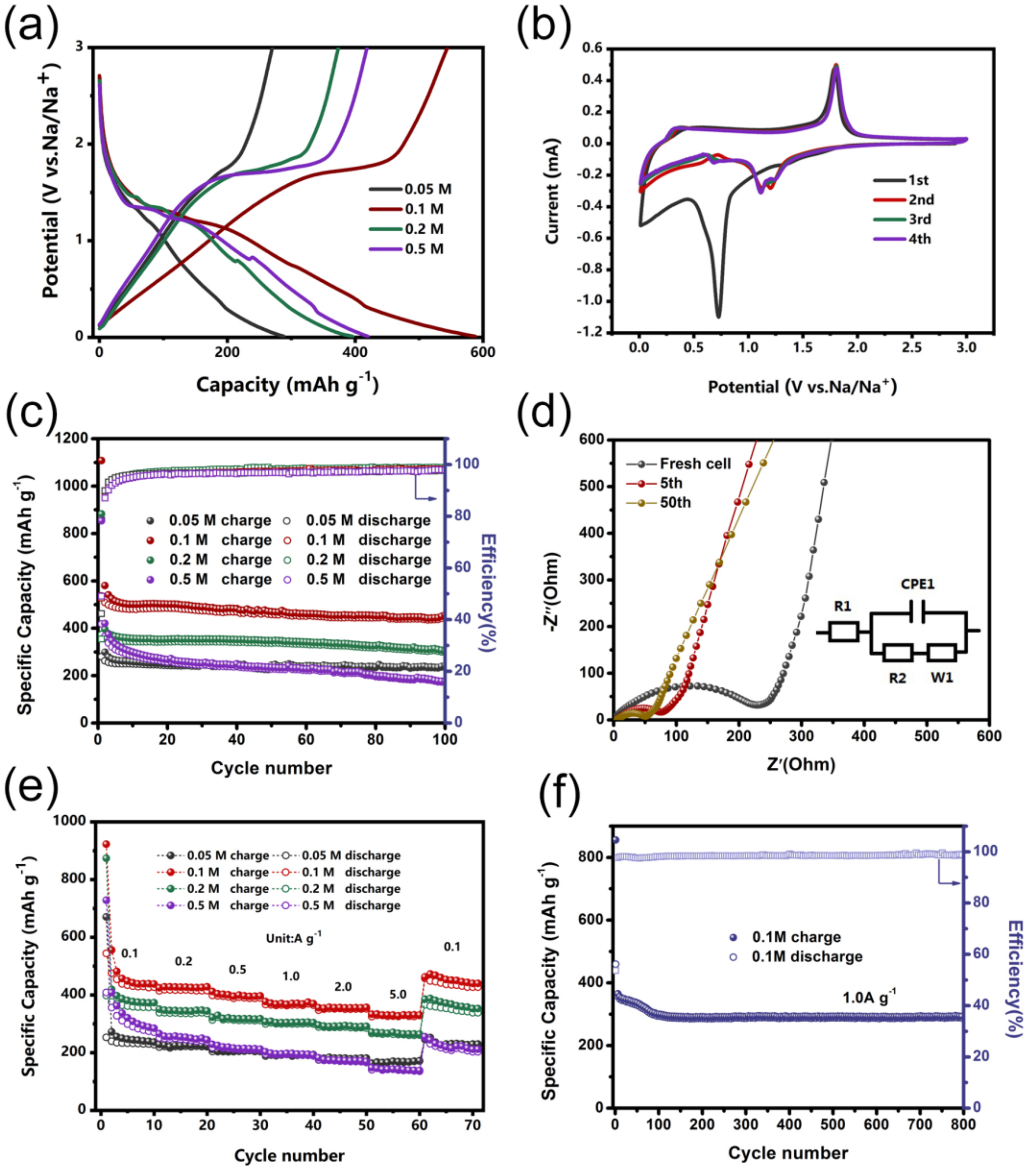

**Figure 6.** Electrochemical performance of CSNC electrode versus $Na/Na^+$. a) Galvanostatic discharge/charge curves of CSNC-0.05, CSNC-0.1, CSNC-0.2 and CSNC-0.5 at the specific current density of 0.1 A $g^{-1}$. b) CV curves of CSNC-0.1 at a sweep rate of 0.1 mV $s^{-1}$. c) Cycle performance at the current density of CSNC-0.05, CSNC-0.1, CSNC-0.2 and CSNC-0.5 at 0.1 A $g^{-1}$. d) The 1$^{st}$, 5$^{th}$ and 50$^{th}$ cycles EIS of CSNC-0.1, fitting circuit model is depicted in inset. e) Rate performances of CSNC-0.05, CSNC-0.1, CSNC-0.2 and CSNC-0.5. e) Cycle performance of CSNC-0.1 at the current density of 1.0 A $g^{-1}$.

The electrochemical performance of CSNC anode for SIBs was evaluated and summarized in Figure 6. Figure 6a exhibits the constant current charge and discharge curves for the second cycle of CSNC at a current density of 0.1 A $g^{-1}$. Each CSNC has obvious charge and discharge platforms, which are consistent with the reduction/oxidation peaks position of the second cycle CV curves shown in Figure S7a. The reduction/oxidation peaks at 1.84 and 1.22 V can well accord with the characteristic peaks of $Co_3Se_4$, and the electrochemical equation can be summarized as

[12]:

$$Co_3Se_4 + 8Na + 8e^- \longleftrightarrow 3Co + 4Na_2Se \tag{4}$$

The reactants and products that can be completely converted to each other means the electrochemical process will be stable in the SIB. The discharge capacity for the second cycle of CSNC-0.05, CSNC-0.1, CSNC-0.2 and CSNC-0.5 respectively reach 288.4, 588.2, 396.1 and 420.1 mAh $g^{-1}$. It is obvious that the capacity of CSNC-0.1 is significantly higher than that of other samples. So the initial four CV cycles of CSNC-0.1 were performed at a sweep rate of 0.1 mV $s^{-1}$ and the results are displayed in Figure 6b. It can be seen that the capacity of the first cycle is greatly reduced, and the potential platform is shifted from 0.72 V to 1.22 V, due to the formation of the SEI film. The capacity storage of shifted potential platform will reduce, but the formation of dendrite can be drastically avoided to keep the SIB safety. Furthermore, the 2-4 cycles of CV curves are almost coincident, which can preliminarily explain the stability of the CSNC in SIB. Figure 6c and Figure S8a display the cycling performance of CSNCs and N-CN at the current density of 0.1 A $g^{-1}$. After 100 cycles, the capacity and the retention rate of N-CN and CSNC against the second cycle are N-CN (229.7 mAh $g^{-1}$, 92.4%), CSNC-0.05 (238.7 mAh $g^{-1}$, 90.9%), CSNC-0.1 (448.7 mAh $g^{-1}$, 88.1%), CSNC-0.2 (298.6 mAh $g^{-1}$, 83.8%) and CSNC-0.5 (172.2 mAh $g^{-1}$, 47.1%), respectively. Of which N-CN shows the best cycle stability, due to the good stable interconnected porous structure. After repeated cycles, the structure does not change substantially, resulting in a stable structure and a good cycle stability of CSNC samples as well. On the other hand, the transition metal selenide possesses higher theoretical capacity, compared with carbon materials. Therefore, the combination of appropriate proportion of $Co_3Se_4$ on N-CN is beneficial to increase the composite capacity. However, with the increase of $Co_3Se_4$ loading, the cycle stability of $Co_3Se_4$@N-CN decreases. This is because that as the proportion of $Co_3Se_4$ nanoparticle increases, $Co_3Se_4$ will cause more volume expansion when the insertion and depletion of $Na^+$ in the process of the charge and discharge reaction in the battery. And the active material will fall off from the N-CN, then the capacity decreases. The best performance battery material was obtained by adjusting the ratio of $Co_3Se_4$. Among the CSNC, as the excellent simple, CSNC-0.1 possesses the highest specific capacity (448.7 mAh $g^{-1}$) with a good cycling stability. The AC impedance diagram of CSNC before cycling is revealed in Figure S7b, which proves that the higher ratio of $Co_3Se_4$, the worse conductivity of the CSNC, overmuch $Co_3Se_4$ will block the charge transfer. It can be well fitted by a classic SIBs equivalent circuit (the inset of Figure S7b). In order to study the subsequent reaction process in the SIB, the EIS of $5^{th}$ and $50^{th}$ cycles of CSNC-0.1 for SIBs are also shown in the Figure 6d. There was no change in the shapes of the $5^{th}$ and $50^{th}$ cycles EIS curves and the equivalent circuit was exactly the same with that of the fresh cell. It proves that there is no new interfacial resistance during the charge and discharge reaction in the SIB. Figure 6e exhibits a comparison of the rate performance of a series of CSNC. Among them, the CSNC-0.1 has the highest capacity and an excellent rate performance, 427.4 mAh $g^{-1}$ (100%) at 0.1 A $g^{-1}$, 415.6 mAh $g^{-1}$ (97.2%) at 0.2 A $g^{-1}$, 389.9 mAh $g^{-1}$ (91.2%) at 0.5 A $g^{-1}$, 365.1 mAh $g^{-1}$ (85.4%) at 1.0 A $g^{-1}$, 352.3 mAh $g^{-1}$ (82.4%) at 2.0 A $g^{-1}$, 328.9

mAh $g^{-1}$ (77.2%) at 5.0 A $g^{-1}$, 431.1 mAh $g^{-1}$ (100.8%) at 0.1 A $g^{-1}$. The rate performance of N-CN is depicted in Figure S8b. The good rate performance of N-CN provides the basis for the excellent rate performance of CSNC-0.1. In view of the high capacity of CSNC-0.1 with good cycle stability and excellent rate performance, we carried out a high current cycle at the current density of 1.0 A $g^{-1}$ to test the performance of CSNC-0.1 and the results are depicted in Figure 6f. In the first 100 cycles of CSNC-0.1, the capacity decreased from 361.1 to 294.8 mAh $g^{-1}$. It may be that at a large current density, $Na^+$ rapidly intercalate and desorb the active material, causes the $Co_3Se_4$ expands sharply and partially falls off from the N-CN. After that, the capacity of CSNC-0.1 tends to be stable, and after 800 cycles, there is still a capacity of 288.4 mAh $g^{-1}$ reserved.

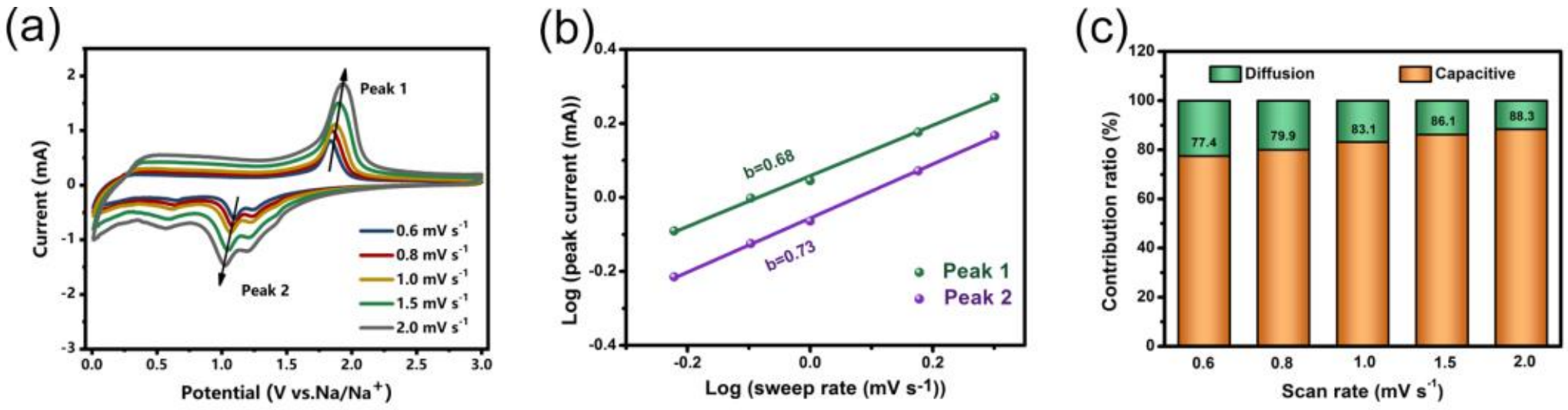

**Figure 7.** a) CV curves at different scan rates ranging from 0.2 to 2 mV $s^{-1}$ . b) Log (Peak current) versus log (scan rate) at different scan rates. c) Contribution ratio of capacitive and diffusion-controlled capacities at different scan rates.

The kinetic properties of CSNC-0.1 in SIB were also studied and displayed in Figure 7. Figure 7a shows the CV curves of CSNC-0.1 at different scan rates (0.6, 0.8, 1.0, 1.5 and 2.0 mV $s^{-1}$). The shapes of different curves are similar accompany the scan rates increase. As depicted in Figure 7b, the $b$ value is calculated to be 0.68 of the anodic process (peak 1 of SIB), and 0.73 of the cathodic process (peak 2 of SIB). It reveals that the energy storage of SIB includes capacitive behavior and diffusion-controlled processes. The contributions of the diffusion-controlled and capacitive contributions of SIB were calculated and shown in Figure 7c. A high capacitive contribution increases from 77.4% to 88.3% of SIB, as the scan rate changes from 0.2 to 2 mV $s^{-1}$, revealing the excellent rate performance of SIB [42].

## Conclusion

In conclusion, we report a 3D interconnected carbon network prepared by an in-situ template method, and $Co_3Se_4$@N-CN was further synthesized as the anode of high performance lithium ion and sodium ion batteries. N-CN possesses several advantages: (1) N-CN has a high specific surface and abundant pore size, which can provide a large number of ion transport channels to increase the reaction speed. (2) The unique stable three-dimensional structure can reduce the capacity loss by protecting the active material from violently volume expansion. (3) N-doping can make N-CN more easily to adsorb $Co^{2+}$, and help N-CN to provide tantalum capacity during the reaction. $Co_3Se_4$ has a higher theoretical capacity than carbon materials, so we prepared CSNC to enhance the overall electrochemical performance of the composite by synergistic effects. In this experiment, we optimized the mass ratio of $Co_3Se_4$ in CSNC by adjusting the concentration of $Co^{2+}$ soaked with N-CN to obtain the best ratio. In the case of lithium ion battery, CSNC-0.1 possesses a constant increased capacity of 1313.5 mA $g^{-1}$ at a current density of 0.1 A $g^{-1}$ after 100 cycles, and the phenomenon of increased capacity is fully explained above. The CSNC-0.1 shows a capacity of 835.6 mAh $g^{-1}$ at a current density of 1.0 A $g^{-1}$ after 500 cycles. In addition, in the sodium ion battery test, CSNC-0.1 achieved a capacity of 448.7 mAh $g^{-1}$ at 0.1 A $g^{-1}$ after 100 cycles, and the reversible capacity reached 328.9 mAh $g^{-1}$ at 5 A $g^{-1}$, which is 77.2% of the capacity at 0.1 A $g^{-1}$. Compared with the transition metal selenide, cobalt sulfide and cobalt oxide anodes for LIBs or SIBs, the CSNC shows much better rate and cycling performances, as listed in Table S1. More importantly, this strategy paves the way for the application of the compounds of carbon materials and transition metal selenide in other fields.

# Supporting Information

**Highly Conductive $Co_3Se_4$ Embedded in N-doped 3D Interconnected Carbonaceous Network for Enhanced Lithium and Sodium Storage**

*Bingke Liu,*[a] *Junming Cao,*[a] *Junzhi Li,*[a] *La Li,* [a*] *Duo Chen,*[a] *Siqi Zhang,*[c] *Dong Cai,*[a] *and Wei Han,*[a,b*]

[a]Sino-Russian international joint laboratory for clean energy and energy conversion technology, College of Physics, Jilin University, Changchun City 130012, P. R. China.
E-mail: lali910217@gmail.com
[b]International Center of Future Science, Jilin University, Changchun City 130012, P. R. China.
E-mail: whan@jlu.edu.cn
[c]Key Laboratory of Physics and Technology for Advanced Batteries (Ministry of Education), College of Physics, Jilin University, Changchun 130012, P. R. China.

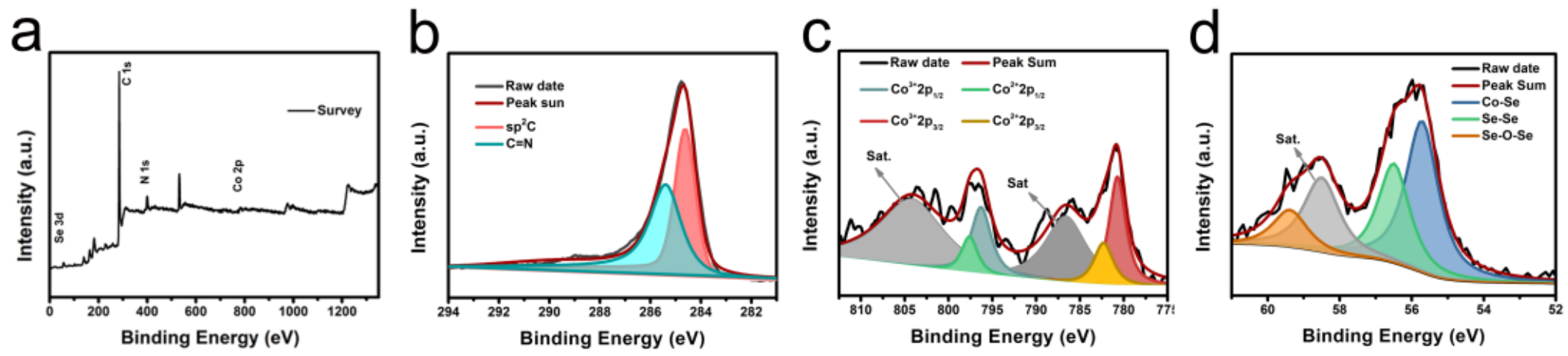

**Figure S1**. XPS spectra of the $Co_3Se_4$@N-CN : a) survey spectrum, b) C 1s, c) Co 2p and (d) Se 3d.

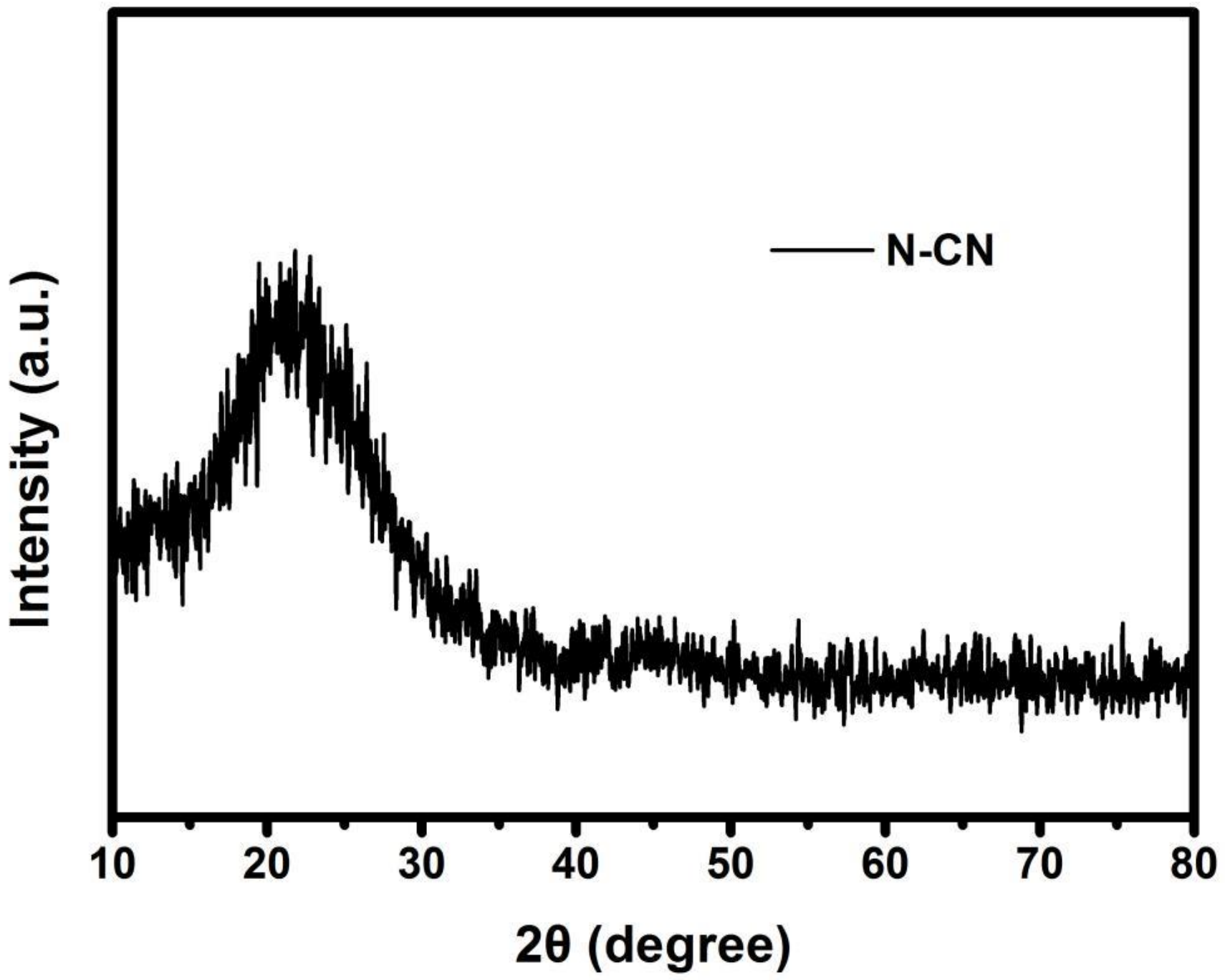

**Figure S2**. XRD patterns of N-CN

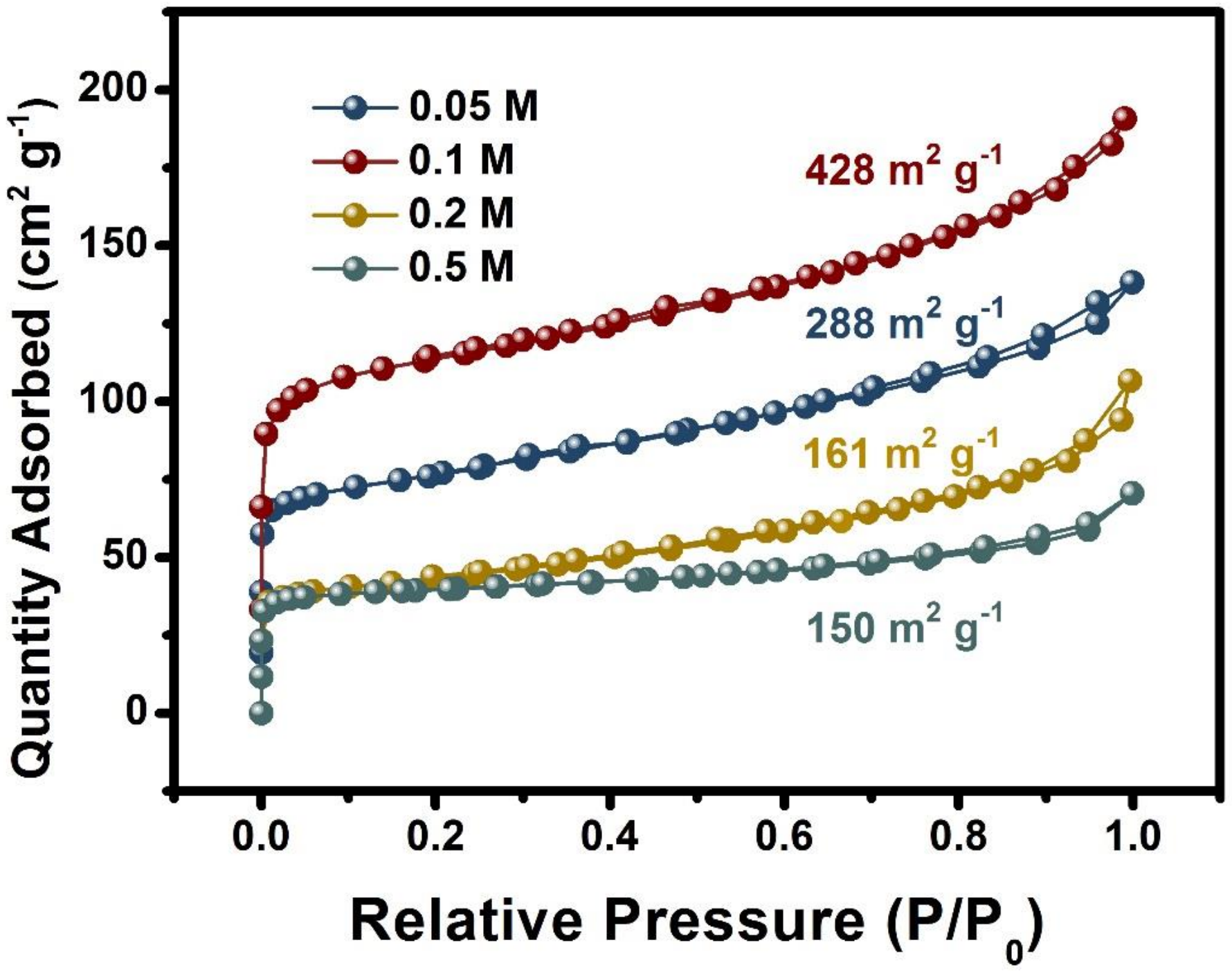

**Figure S3.** The $N_2$ adsorption-desorption isotherms of CSNC-0.05, CSNC-0.1, CSNC-0.2, and CSNC-0.5.

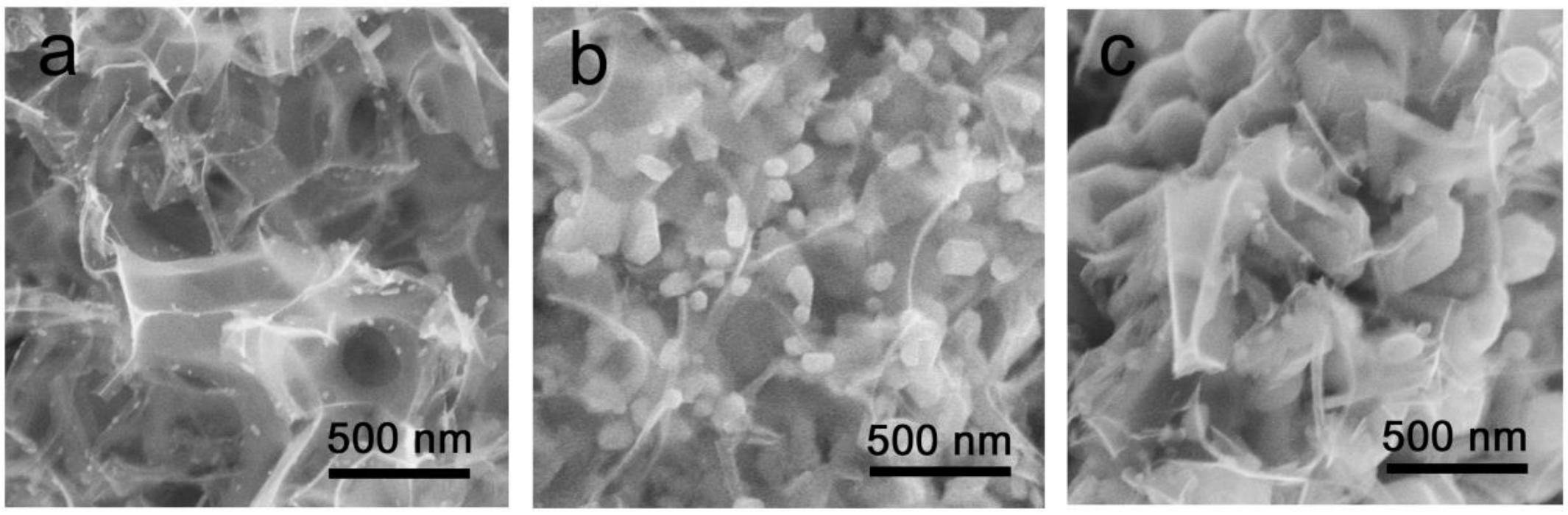

**Figure S4.** SEM of a) CSNC-0.05. b) CSNC-0.2. c) CSNC-0.5.

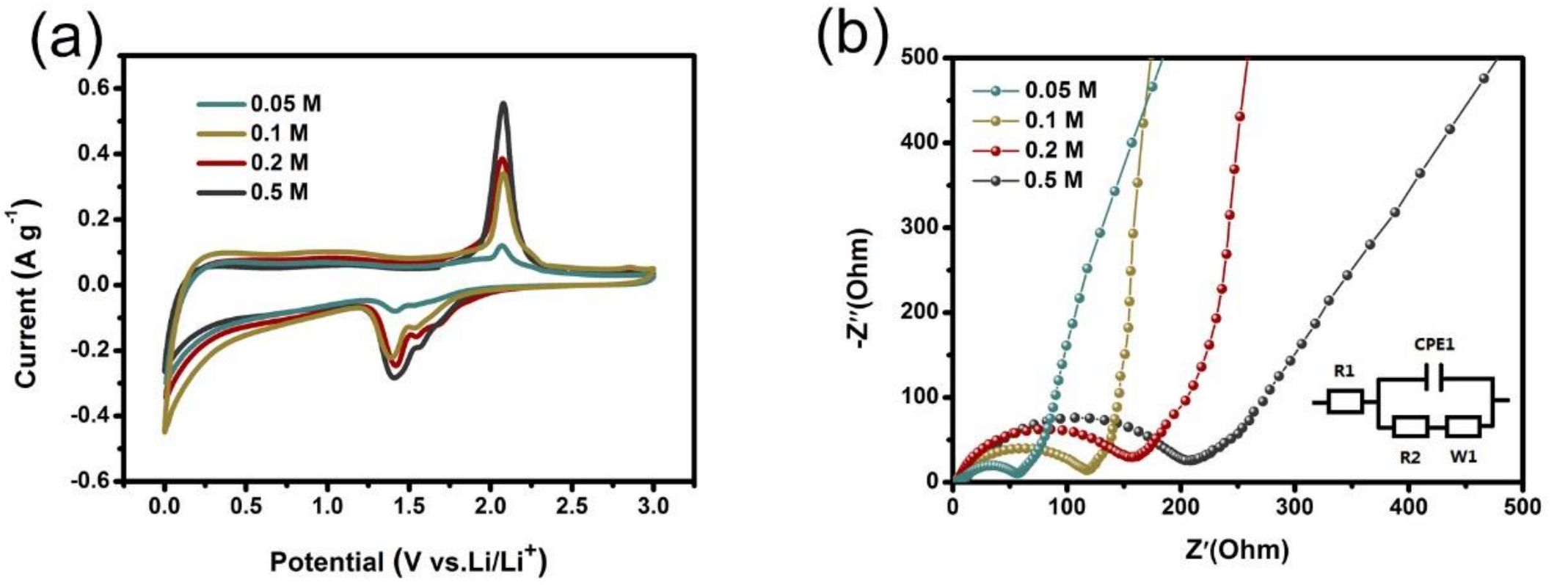

**Figure S5**. a) CV curves at a sweep rate of 0.1 mV $s^{-1}$ and b) fresh cell EIS versus $Li/Li^{+}$ of NSCN-0.05, NSCN-0.1, NSCN-0.2 and NSCN-0.5.

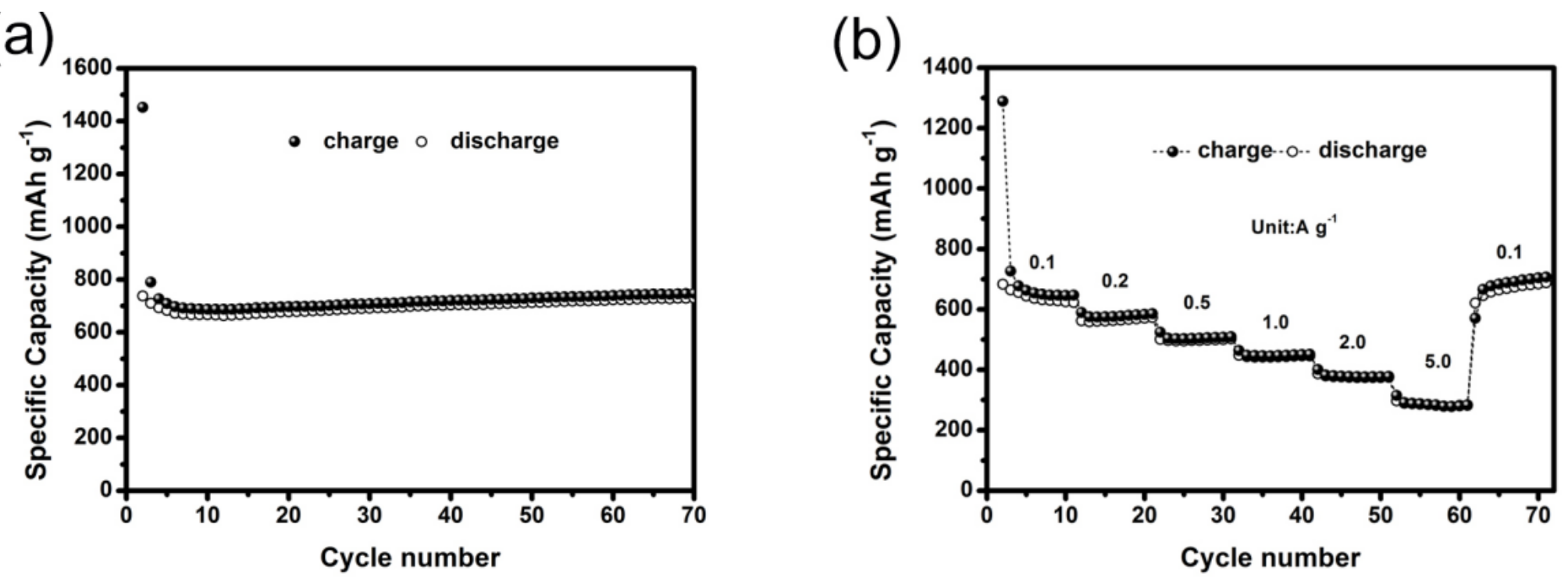

**Figure S6.** a) Cycle performance of N-CN at 0.1 A $g^{-1}$, b) rate performance of N-CN versus $Li/Li^{+}$.

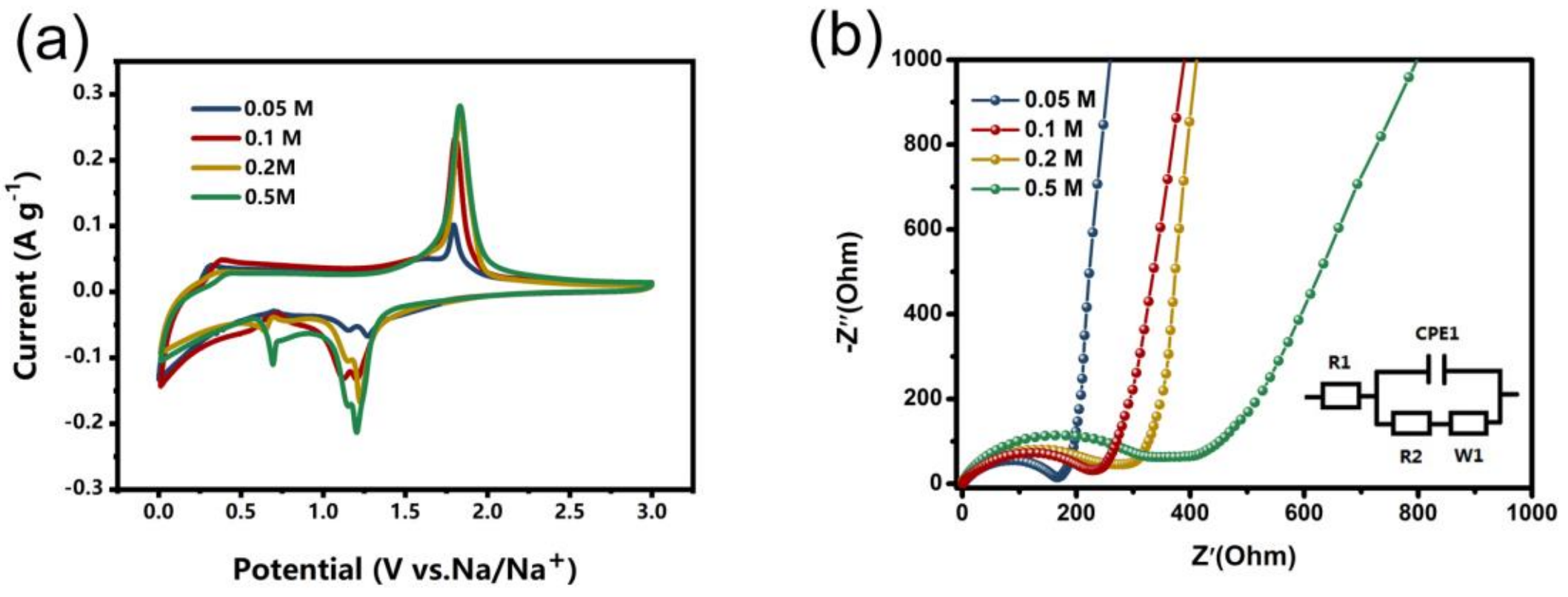

**Figure S7**. a) CV curves at a sweep rate of 0.1 mV $s^{-1}$. b) EIS results of CSNC-0.05, CSNC-0.1, CSNC-0.2 and CSNC-0.5, fitting circuit model is depicted in inset.

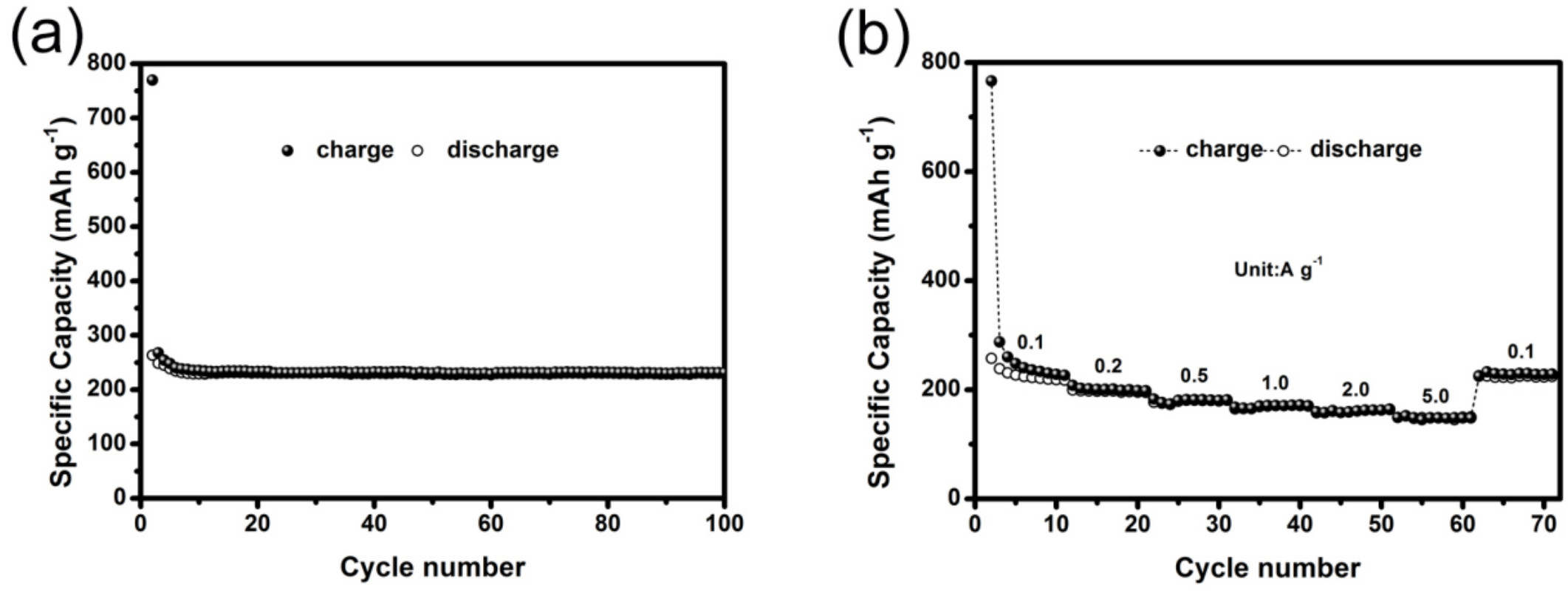

**Figure S8.** a) Cycle performance of N-CN at 0.1 A $g^{-1}$, b) rate performance of N-CN versus $Na/Na^{+}$.

**Table S1.** Comparison of electrochemical performance of transition metal selenides, cobalt sulfide and cobalt oxide anodes for LIBs or SIBs.

| Materials | LIBs (mAh $g^{-1}$) | | SIBs (mAh $g^{-1}$) | | Year/Ref |
|---|---|---|---|---|---|
| | Cycling | Rate properties | Cycling | Rate properties | |
| $Co_3Se_4$@N-CN | **1313.5** after **100** cycles at **0.1** A $g^{-1}$ | **760.6** at **1.0** A $g^{-1}$ **544.5** at **5.0** A $g^{-1}$ | **448.7** after **100** cycles at **0.1** A $g^{-1}$ | **352.3** at **1.0** A $g^{-1}$ **328.9** at **5.0** A $g^{-1}$ | This work |
| CoSe@C | **675.0** after **100** cycles at **0.2** A $g^{-1}$ | **590.0** at **1.0** A $g^{-1}$ **198.6** at **5.0** A $g^{-1}$ | **341.0** after **100** cycles at **0.1** A $g^{-1}$ | **278.9** at **1.0** A $g^{-1}$ **207.7** at **4.0** A $g^{-1}$ | 2017 [1] |
| NiSe@C | **428.0** after **50** cycles at **0.1** A $g^{-1}$ | **435.0** at **0.1** A $g^{-1}$ **299.0** at **0.5** A $g^{-1}$ | **280.0** after **50** cycles at **0.1** A $g^{-1}$ | **263.0** at **0.1** A $g^{-1}$ **186.0** at **0.5** A $g^{-1}$ | 2016 [2] |
| MoSe@N-C | **1219** after **150** cycles at **1.0** A $g^{-1}$ | **807.0** at **2.0** A $g^{-1}$ **448.0** at **6.0** A $g^{-1}$ | | | 2018 [3] |
| CoSe/Co@NC | **630.0** after **100** cycles at **0.2** A $g^{-1}$ | **328.0** at **1.0** A $g^{-1}$ **185.0** at **5.0** A $g^{-1}$ | | | 2018 [4] |
| $SnSe_2$@rGO | **745.7** after **500** cycles at **0.2** A $g^{-1}$ | **532.1** at **1.0** A $g^{-1}$ **325.4** at **5.0** A $g^{-1}$ | | | 2018 [5] |
| $CoSe_2$@C | | | **430.0** after **400** cycles at **0.2** A $g^{-1}$ | **383.0** at **0.8** A $g^{-1}$ **278.0** at **3.2** A $g^{-1}$ | 2019 [6] |
| $CoSe_2$@C | | | **371.8** after **500** cycles at **0.2** A $g^{-1}$ | **307.0** at **1.0** A $g^{-1}$ **299.2** at **2.0** A $g^{-1}$ | 2018 [7] |
| $Fe_7Se_8$@NC | | | **337.0** after **100** cycles at **0.5** A $g^{-1}$ | **293.0** at **1.0** A $g^{-1}$ **172.0** at **5.0** A $g^{-1}$ | 2018 [8] |
| $MoSe_2$/N,P-rGO | | | **378.0** after **1000** cycles at **0.5** A $g^{-1}$ | **351.0** at **1.0** A $g^{-1}$ **310.0** at **5.0** A $g^{-1}$ | 2017 [9] |
| $NiSe_2$@rGO | | | **346.0** after **1000** cycles at **1.0** A $g^{-1}$ | **366.0** at **1.0** A $g^{-1}$ **318.0** at **5.0** A $g^{-1}$ | 2017 [10] |
| CoSe@C | | | **299.0** after **100** cycles at **0.1** A $g^{-1}$ | **308.0** at **0.5** A $g^{-1}$ **241.0** at **5.0** A $g^{-1}$ | 2016 [11] |
| $Co_9S_8$@C | **768.0** after **100** cycles at **0.1** A $g^{-1}$ | **447.0** at **1.0** A $g^{-1}$ **287**.0 at **2.0** A g-1 | **421.0** after **50** cycles at **0.1** A $g^{-1}$ | **288.0** at **1.0** A $g^{-1}$ **252.0** at **2.0** A $g^{-1}$ | 2019 [12] |
| $Co_3O_4$/N-C | **507.0** after **1000** cycles at **2.0** A $g^{-1}$ | **753.0** at **1.0** A $g^{-1}$ **529.0** at **5.0** A $g^{-1}$ | **368.0** after **50** cycles at **0.1** A $g^{-1}$ | **385.0** at **1.0** A $g^{-1}$ **326.0** at **5.0** A $g^{-1}$ | 2017 [13] |
| $Co_3O_4$@C | | | **300.0** after **100** cycles at **0.05** A $g^{-1}$ | **285.5** at **1.0** A $g^{-1}$ **225.1** at **2.0** A $g^{-1}$ | 2017 [14] |
| $Co_3S_4$@PANI | | | **252.5** after **100** cycles at **0.2** A $g^{-1}$ | **218.6** at **1.0** A $g^{-1}$ **184.1** at **4.0** A $g^{-1}$ | 2016 [15] |
| $Co_3O_4$@NC | | | **373.0** after **60** cycles at **0.2** A $g^{-1}$ | **317.0** at **0.4** A $g^{-1}$ **263.0** at **1.0** A $g^{-1}$ | 2016 [16] |